\documentclass[prc,showpacs,superscriptaddress,twocolumn]{revtex4}

\usepackage{amsmath,bm}
\usepackage{graphicx}
\usepackage{epstopdf}
\usepackage{subfigure}
\usepackage{epsfig}
\usepackage{amsmath,amssymb,amsfonts}
\usepackage{color}
\usepackage[utf8]{inputenc}
\usepackage{verbatim}
\usepackage{array}

\begin{document}

\newcommand{\vk}{{\vec k}}
\newcommand{\vK}{{\vec K}}
\newcommand{\vb}{{\vec b}}
\newcommand{{\vp}}{{\vec p}}
\newcommand{{\vq}}{{\vec q}}
\newcommand{\vQ}{{\vec Q}}
\newcommand{\vx}{{\vec x}}
\newcommand{\beq}{\begin{equation}}
\newcommand{\eeq}{\end{equation}}
\newcommand{\half}{{\textstyle \frac{1}{2}}}
\newcommand{\gton}{\stackrel{>}{\sim}}
\newcommand{\lton}{\mathrel{\lower.9ex \hbox{$\stackrel{\displaystyle<}{\sim}$}}}
\newcommand{\ee}{\end{equation}}
\newcommand{\ben}{\begin{enumerate}}
\newcommand{\een}{\end{enumerate}}
\newcommand{\bit}{\begin{itemize}}
\newcommand{\eit}{\end{itemize}}
\newcommand{\bc}{\begin{center}}
\newcommand{\ec}{\end{center}}
\newcommand{\bea}{\begin{eqnarray}}
\newcommand{\eea}{\end{eqnarray}}

\newcommand{\beqar}{\begin{eqnarray}}
\newcommand{\eeqar}[1]{\label{#1} \end{eqnarray}}
\newcommand{\pleft}{\stackrel{\leftarrow}{\partial}}
\newcommand{\pright}{\stackrel{\rightarrow}{\partial}}

\newcommand{\eq}[1]{Eq.~(\ref{#1})}
\newcommand{\fig}[1]{Fig.~\ref{#1}}
\newcommand{\eff}{ef\!f}
\newcommand{\alphas}{\alpha_s}

\renewcommand{\topfraction}{0.85}
\renewcommand{\textfraction}{0.1}

\renewcommand{\floatpagefraction}{0.75}

\title{$p_{T}$ dispersion of inclusive jets in high-energy nuclear collisions}

\date{\today  \hspace{1ex}}

\author{Shi-Yong Chen}
\affiliation{Huanggang Normal University, Huanggang 438000, China}
\affiliation{Key Laboratory of Quark \& Lepton Physics (MOE) and Institute of Particle Physics,
 Central China Normal University, Wuhan 430079, China}

\author{Jun Yan}
\affiliation{Key Laboratory of Quark \& Lepton Physics (MOE) and Institute of Particle Physics,
 Central China Normal University, Wuhan 430079, China}

\author{Wei Dai}
\affiliation{School of Mathematics and Physics, China University of Geosciences, Wuhan 430074, China}

\author{Ben-Wei Zhang\footnote{bwzhang@mail.ccnu.edu.cn}}
\affiliation{Key Laboratory of Quark \& Lepton Physics (MOE) and Institute of Particle Physics,
 Central China Normal University, Wuhan 430079, China}
\affiliation{Guangdong Provincial Key Laboratory of Nuclear Science,Institute of Quantum Matter, South China Normal University, Guangzhou 510006, China}

\author{Enke Wang}
\affiliation{Guangdong Provincial Key Laboratory of Nuclear Science,Institute of Quantum Matter, South China Normal University, Guangzhou 510006, China}
\affiliation{Key Laboratory of Quark \& Lepton Physics (MOE) and Institute of Particle Physics,
 Central China Normal University, Wuhan 430079, China}

\begin{abstract}
In this study, we investigate the impact of jet quenching on the $p_{T}$ dispersion($p_{T}D$) of inclusive jets ($R=0.2$) in central Pb+Pb (0-10\%) collisions at $\sqrt{s}=2.76$~TeV.
The partonic spectrum in the initial hard scattering of elementary collisions is obtained by an event generator POWHEG+PYTHIA, which matches the next-to-leading order (NLO) matrix elements with parton showering, and the energy loss of a fast parton traversing through hot/dense QCD medium is calculated using Monte Carlo simulation within the Higher-Twist formalism of jet quenching in heavy-ion collisions.
We present model calculations of the normalized $p_{T}D$ distributions of inclusive jets in p+p and central Pb+Pb collisions at $\sqrt{s}=2.76$~TeV, which offer good descriptions of ALICE measurements. It is shown that the $p_{T}D$ distributions of inclusive jets in central Pb+Pb collisions shift significantly to a higher $p_{T}D$ region relative to those in p+p collisions. Thus the nuclear modification ratio of the $p_{T}D$ distributions of inclusive jets is smaller than unity in the small $p_{T}D$ region, and larger than one in the large $p_{T}D$ region. This behavior is caused by a more uneven $p_T$ distribution of jet constituents as well as the fraction alteration of quark/gluon initiated jets in heavy-ion collisions. The difference in $p_{T}D$ distribution between groomed and ungroomed jets in Pb+Pb collisions is also discussed.

\end{abstract}


\maketitle

\section{Introduction}
\label{sec:Intro}

Energetic partons created in the early stage of heavy-ion collisions (HIC) may suffer energy loss owing to their interacting with the quark-gluon plasma (QGP), a novel state of matter with deconfined quarks and gluons under extreme high temperature and energy density. This
phenomenon is referred as jet quenching~\cite{Wang:1991xy,Gyulassy:2003mc,Qin:2015srf}, which could provide powerful tools to study the creation and properties of the QCD medium.  In the last decade the investigations of jet quenching have been extended from the leading hadron productions suppression~\cite{Khachatryan:2016odn, Acharya:2018qsh, Aad:2015wga, Burke:2013yra,Chen:2010te,Chen:2011vt,Liu:2015vna,Dai:2015dxa,Dai:2017piq,Dai:2017tuy,Ma:2018swx,Xie:2019oxg,Zhang:2022fau} to medium modifications of a wealth of reconstructed jet observables, such as inclusive jets productions, di-jets asymmetry, correlations of gauge bosons and jets , as well as heavy flavor jets~\cite{Vitev:2008rz,Vitev:2009rd,Aad:2010bu,Chatrchyan:2011sx,Chatrchyan:2012gt,Aad:2014bxa,Chatrchyan:2012gw,Chatrchyan:2013kwa, Aad:2014wha,Sirunyan:2017jic, CasalderreySolana:2010eh, Sirunyan:2018ncy,Young:2011qx,He:2011pd,ColemanSmith:2012vr,Neufeld:2010fj,Zapp:2012ak, Dai:2012am, Ma:2013pha, Senzel:2013dta, Casalderrey-Solana:2014bpa,Milhano:2015mng,Chang:2016gjp,Majumder:2014gda, Chen:2016cof, Chien:2016led, Apolinario:2017qay,Connors:2017ptx,Zhang:2018urd,Dai:2018mhw,Luo:2018pto,Chang:2019sae,Wang:2019xey,Chen:2019gqo,Chen:2020kex,Wang:2020qwe,Yan:2020zrz,Wang:2020ukj,Zhang:2021sua}.
A full reconstructed jet is a collimated spray of hadrons created in $e^+e^-$ collisions, p+p reactions as well as nucleus-nucleus collisions with large momentum transfer, and the existence of QCD medium
should naturally modify the yields and the internal structures of full jets, and thus the medium modifications of jet observables could be used to tomography of QGP formed in HIC.

The nuclear modifications of jet substructure have received a growing attention in the heavy-ion community.
One interesting jet substructure is jet $p_{T}$ dispersion($p_{T}D$), which characterizes the fragmentation of a jet~\cite{Giele:1997hd,Acharya:2018uvf,KunnawalkamElayavalli:2017hxo,Agafonova:2019tqe,Wan:2018zpq}.  The nuclear modification of jet $p_{T}D$ distribution may improve our understanding of jet-medium interaction, and put new insight on how jet substructure is resolved by QCD medium. Recently, ALICE Collaboration has measured $p_{T}D$ distributions for small-radius ($R=0.2$) jets in heavy-ion collisions~\cite{Acharya:2018uvf}, which further facilitates the studies of $p_{T}D$ in distributions HIC, since the theoretical calculations could be confronted with the data directly to infer some crucial information of jet propagation in the QCD medium.

In this paper, we present our study on the normalized $p_{T}D$ distributions for inclusive jets with jet radius $R=0.2$ both in p+p and central Pb+Pb collisions at $\sqrt{s_{NN}}=2.76$~TeV. We employ POWHEG+PYTHIA~\cite{Alioli:2010xa,Alioli:2010qp,Buckley:2016bhy},  a Monte Carlo model matching NLO matrix elements with parton shower (PS), including hadronization process to obtain the solid baseline results of jet $p_{T}D$ in p+p collisions, which are then served as input to simulate parton energy loss within the higher-twist approach~\cite{Guo:2000nz,Zhang:2003yn,Zhang:2003wk,Majumder:2009ge} to compute the $p_{T}D$ distribution in heavy-ion collisions. Our model calculations of $p_{T}D$ distribution for inclusive jets could provide satisfactory descriptions of ALICE data both in p+p and Pb+Pb collisions, where we observe a shift of $p_{T}D$ distribution toward higher values in Pb+Pb collisions relative to that in p+p. We further make a comprehensive understanding of the distinct feature between quark and gluon initiated jets, and the nuclear modification ratio $p_{T}D$ distribution. We find $p_{T}D$ can be analytical expressed as the standard deviation and the multiplicity of jet constitutes. After jet quenching, more jet constituents lied further from the mean value of $p_{T}$.

The remainder of this paper is organized as follows. In Sec.~\ref{sec:framework} we will introduce the framework used to calculate the normalized $p_{T}D$ distributions in both p+p and central Pb+Pb collisions.  Our numerical results and detailed discussions of the medium modifications of $p_{T}D$ distributions for groomed and ungroomed jets are presented in Section~\ref{sec:results}. In Sec.~\ref{sec:summary} we will give a summary.

\section{Analysis framework}
\label{sec:framework}
We study a jet substructure observable, the $p_{T}D$, which characterizes the second moment of the constituent $p_{T}$ distribution inside a jet~\cite{Giele:1997hd,Acharya:2018uvf}, and is defined
as:
\begin{eqnarray}
p_{T}D=\frac{\sqrt{\sum_{i}p^{2}_{T,i}}}{p_{T,{\rm jet}}}
\label{eq:g}
\end{eqnarray}
where $p_{T,i}$ represents the transverse momentum of $i$th jet ingredient inside the jet with transverse momentum $p_{T,{\rm jet}}$. $p_{T}D$ is connected to
how hard or soft the jet fragmentation is, and whether $p_{T,i}$ distribution is uniform or not. For example, in the extreme case of very few constituents carrying a lion's share of the jet momentum, $p_{T}D$ will be close to unity; while in the case of jets containing a large number
of constituents with soft momentum, $p_{T}D$ may approach to zero. It is noted that jet dispersion is one of
a class of jet substructure observables, the generalized jet angularities~\cite{Larkoski:2014pca,ALICE:2021njq}, which are defined as $\lambda^{\kappa}_{\beta}=\Sigma_{i}z^{\kappa}_{i}\theta^{\beta}_{i}$, where $z_{i}=p_{T,i}/p_{T,jet}$ is the momentum fraction of jet constituents, $\theta_{i}=\Delta{R_{i}}/R$, $\Delta{R_{i}}$ is the opening angle from constituent to jet axis, $\kappa$ and $\beta$ are free parameters. One can see that  $(p_{T}D)^2$ is equal to  $\lambda^{\kappa}_{\beta}(\kappa=2, \beta=0)$.

In this work, we use a Monte Carlo model POWHEG+PYTHIA, which performs next-to-leading order (NLO) matrix elements matched with parton showering~\cite{Alioli:2010xa,Alioli:2010qp,Buckley:2016bhy},
to generate jet productions in p+p collisions.  In our simulation the POWHEG BOX code is utilized~\cite{Alioli:2010xa,Alioli:2010qp}, which provides a computer framework for performing NLO calculations in parton shower Monte Carlo programs in accordance with the POWHEG method~\cite{Frixione:2007vw}. Previous studies have shown that, POWHEG BOX Monte Carlo program matched with parton showering could give nice description of productions and correlations for a variety of processes in p+p collisions, such as di-jet, gauge boson tagged jets, heavy flavour jets etc.~\cite{powheg-box}. We generate the NLO matrix elements for QCD dijet events with POWHEG BOX, and then matched with PYTHIA6 by Perugia 2011 tunes~\cite{Skands:2010ak} to perform parton showering and hadronization~\cite{Sjostrand:2006za}. After that,
Fastjet package~\cite{Cacciari:2008gp} is employed to reconstruct final state hadrons into full jets.

\begin{figure}[!htb]
\centering
\includegraphics[width=9.5cm,height=9.6cm]{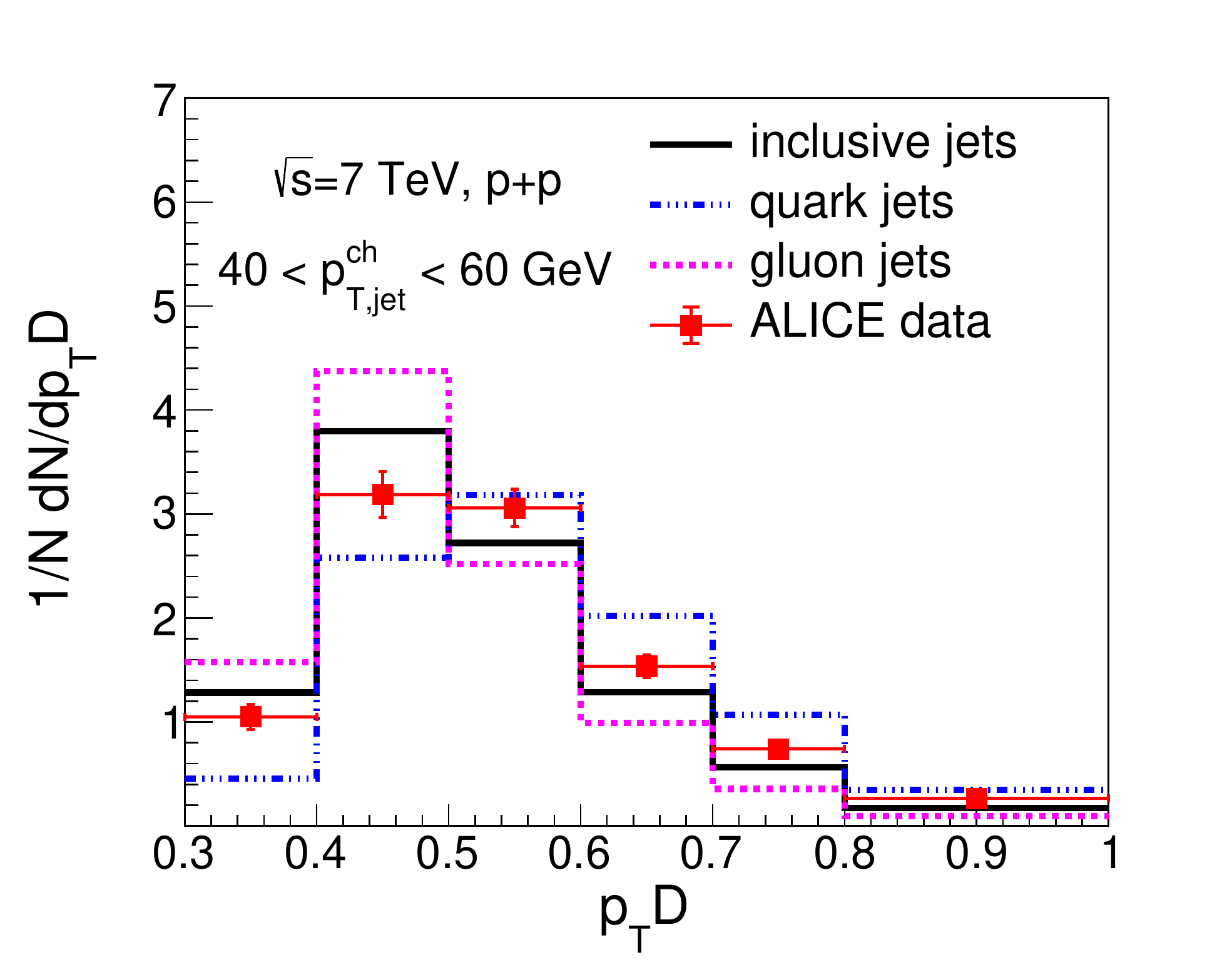} \\
\caption{Normalized $p_{T}D$ distribution of inclusive jets in p+p collisions at
$\sqrt{s}=7$~TeV from POWHEG+PYTHIA calculation as compared with ALICE data~\cite{Acharya:2018uvf}.}
\label{g-ALICE}
\end{figure}

In order to compare with the available experimental data,
we selected events according to the same kinematic
cuts as adopted by the experimental measurements. In ALICE Collaboration data, jets are reconstructed using anti-$k_{T}$
algorithm with radius parameter $R=0.2$ from charged hadrons which are required to have $p_{T}>0.15$~GeV.
Those reconstructed jets are accepted in transverse momentum range of $40 \ {\rm GeV} <p_{T, \rm jet}<60 $~GeV and rapidity range of $\left|\eta_{\rm jet}\right|<0.7$.
Our numerical results of normalized distributions of $p_{T}D$ in p+p collisions at
$\sqrt{s}=7$~TeV and their comparison with ALICE data are shown in Fig.~\ref{g-ALICE}. We can observe that, POWHEG+PYTHIA calculations show well agreements with experimental measurements in p+p collisions in the overall $p_{T}D$ region, which will be served as input for the subsequent study of nuclear modification in HIC. The $p_{T}D$ distributions of quark- and gluon-initiated jets are also plotted in Fig.~\ref{g-ALICE}, respectively.  We find that at the same jet $p_T$, the peak of $p_{T}D$ distribution for gluon jets is located at smaller region relative to that for quark jets. It implies as compared to quark jets, gluon jets favor harder radiation, on average. To make a comprehensive understanding of the distinct feature between quark and gluon jets, we derived the $p_{T}D$ with standard deviation(labeled as $\sigma$) and the multiplicity of jet constitutes(labeled as $n$) in the following.

\begin{figure}[!htb]
\centering
\includegraphics[width=9.5cm,height=9.6cm]{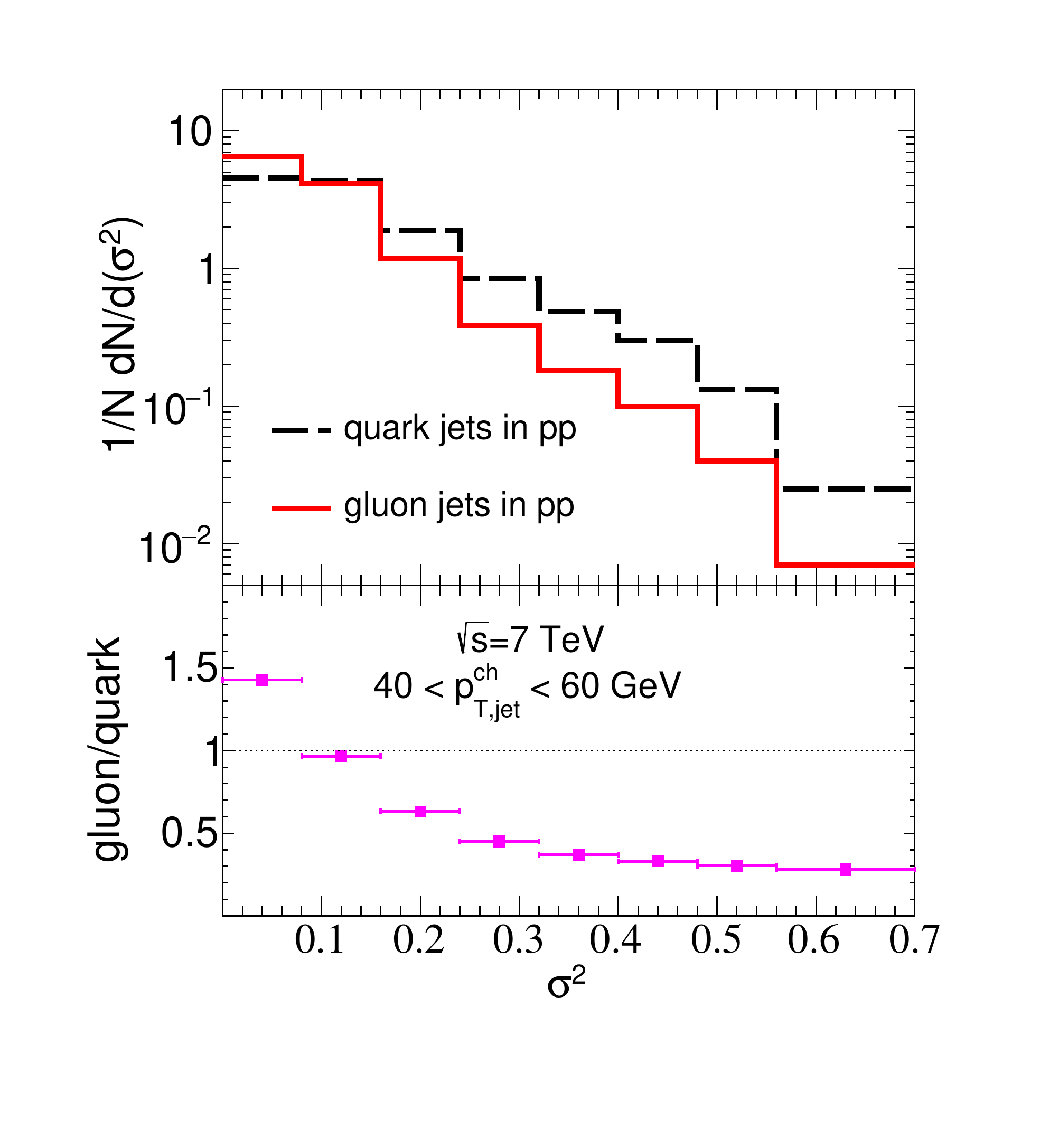} \\
\vspace{-1cm}
\caption{Top panel: normalized $\sigma^2$ distribution of quark- and gluon-initiated jets in p+p collisions at
$\sqrt{s}=7$~TeV from POWHEG+PYTHIA calculation; Bottom panel: ratio of normalized $\sigma^2$ distribution of gluon and quark jets.}
\label{delta2-qg}
\end{figure}

The standard deviation could describe the average degree of a dataset. It tells us, on average, how far each value lies away from the mean value.
A high standard deviation signifies that values are generally far from the mean value, while a low standard deviation means that values are clustered near to the mean value.
Aa a result of that, in our study, the standard deviation of transverse momentum of jet constituents can be written as:

\begin{eqnarray}
\sigma=\frac{\sqrt{\sum_{i}(p_{T,i}-\langle p_{T,i} \rangle)^2}}{p_{T,{\rm jet}}}
\label{eq:g2}
\end{eqnarray}

Then $\sigma^2$ can be expressed as:

\begin{eqnarray}
\sigma^2 &=& \frac{\sum_{i}(p_{T,i}-\langle p_{T,i} \rangle)^2}{(n \cdot \langle p_{T,i} \rangle)^2}  \nonumber  \\
         &=&  \frac{\sum_{i}(p^{2}_{T,i}-2p_{T,i}\langle p_{T,i} \rangle + \langle p_{T,i} \rangle^2)}{(n \cdot \langle p_{T,i} \rangle)^2}  \nonumber \\
\label{eq:g3}
         &=& (p_{T}D)^2-1/n
         \label{eq:sigma}
\end{eqnarray}

Conversely, we have $$(p_{T}D)^2=\sigma^2+1/n \, .$$ Shown in Fig.~\ref{delta2-qg} are the normalized $\sigma^2$ distributions (top) of quark and gluon jets in p+p collisions at $\sqrt{s}=7$~TeV, as well as their gluon/quark ratios (bottom). We observed more gluon jets distributed in lower $p_{T}D$ region compared to quark jets.
It is because gluon jets contain more fragment ingredients. Thus, at the same energy, the value of standard deviation for gluon jets is smaller than that for quark jets.

In heavy-ion collisions, fast partons produced from hard scattering will interact with medium partons and lose their energy. In our calculations, the initial jet shower partons are generated by POWHEG+PYTHIA,
then they are arranged to have initial positions which are sampled from Glauber model~\cite{Alver:2008aq}.
We assume all the partons move in QGP in the same way as classical particles do. The probability for gluon
radiation happens in QGP during each time step $\Delta t$ can be expressed as~\cite{He:2015pra,Cao:2016gvr,Cao:2017hhk,Wang:2019xey}:
\begin{eqnarray}
P_{rad}(t,\Delta t)=1-e^{-\left\langle N(t,\Delta t)\right\rangle} \, .
\label{eq:g4}
\end{eqnarray}

Here $\left\langle N(t,\Delta t)\right\rangle$ is the averaged number of emitted gluons, which can be integrated
from the medium induced radiated gluon spectrum within Higher-Twist(HT) method~\cite{Guo:2000nz,Zhang:2003yn,Zhang:2003wk,Majumder:2009ge}:
\begin{eqnarray}
\frac{dN}{dxdk^{2}_{\perp}dt}=\frac{2\alpha_{s}C_sP(x)\hat{q}}{\pi k^{4}_{\perp}}\sin^2(\frac{t-t_i}{2\tau_f})(\frac{k^2_{\perp}}{k^2_{\perp}+x^2M^2})^4
\label{eq:g5}
\end{eqnarray}

Here $\alpha_{s}$ denotes the strong coupling constant, $x$ is the energy fraction of the radiated gluon, $M$ is the mass of parent parton, and $k_\perp$ is the $p_{T}$ of the radiated gluon.
A lower $p_{T}$ cut-off with $x_{min}=\mu_{D}/E$ of the emitted gluon is applied in our calculations, and $\mu_{D}$ is the Debye screening mass.
 $P(x)$ is the parton splitting function in vacuum, $C_s$ is the Casimir factor for gluons ($C_A$) and quarks ($C_F$). The formation time of the radiated gluons is $\tau_f=2Ex(1-x)/(k^2_\perp+x^2M^2)$.
$\hat{q}$ is jet transport parameter, which is proportional to the local parton distribution density in the QCD medium,
The jet transport parameter $\hat{q}$ is proportional to the local parton density distribution in the QCD medium, and related to the space and time evolution of the medium relative to its initial value $\hat{q}_0$ in the central region when QGP formed, which controls the magnitude of energy loss due to jet-medium interaction.

The number of emitted gluons is sampled from a Poisson distribution during each time step:

 \begin{eqnarray}
P(n_{g},t,\Delta t)=\frac{\left\langle N(t,\Delta t)\right\rangle^{n_{g}}}{n_{g}!}e^{-\left\langle N(t,\Delta t)\right\rangle} \, .
\label{eq:g4}
\end{eqnarray}
In our calculation, $P_{rad}(t,\Delta t)$ would be firstly evaluated to determine whether the radiation happen during $\Delta t$. If accepted, the
Possion distribution $P(n_{g},t,\Delta t)$ is used to sample the number of radiated gluon. At last, the energy fraction ($x$) and transverse momentum ($k_{\perp}$) of the radiated gluon could be
sampled based on the spectrum shown in Eq.5.

To calculate the collisional energy loss of these showered partons\cite{Dai:2018mhw,Wang:2019xey}, a Hard Thermal Loop (HTL) formula has been adopted in this work~\cite{Neufeld:2010xi}:
$\frac{dE^{coll}}{dt}=\frac{\alpha_{s}C_{s}\mu^{2}_{D}}{2}ln\frac{\sqrt{ET}}{\mu_{D}}$.
The space time evolution of bulk medium is given by
the smooth iEBE-VISHNU hydrodynamical code~\cite{Shen:2014vra}.
Jet partons stop their propagation in QGP medium when local temperature fall below $T_c=165$~MeV.
After all the partons escape from QGP, PYQUEN method is used to carry out the hadronization process~\cite{Lokhtin:2000wm,Lokhtin:2005px}. In the model, the radiated gluons are rearranged in the same string as their parent partons,
and these partons could fragment into hadrons
by standard PYTHIA hadronization procedure.

\label{sec:results}

\begin{figure}[!htb]
\centering
\includegraphics[width=9.5cm,height=9.6cm]{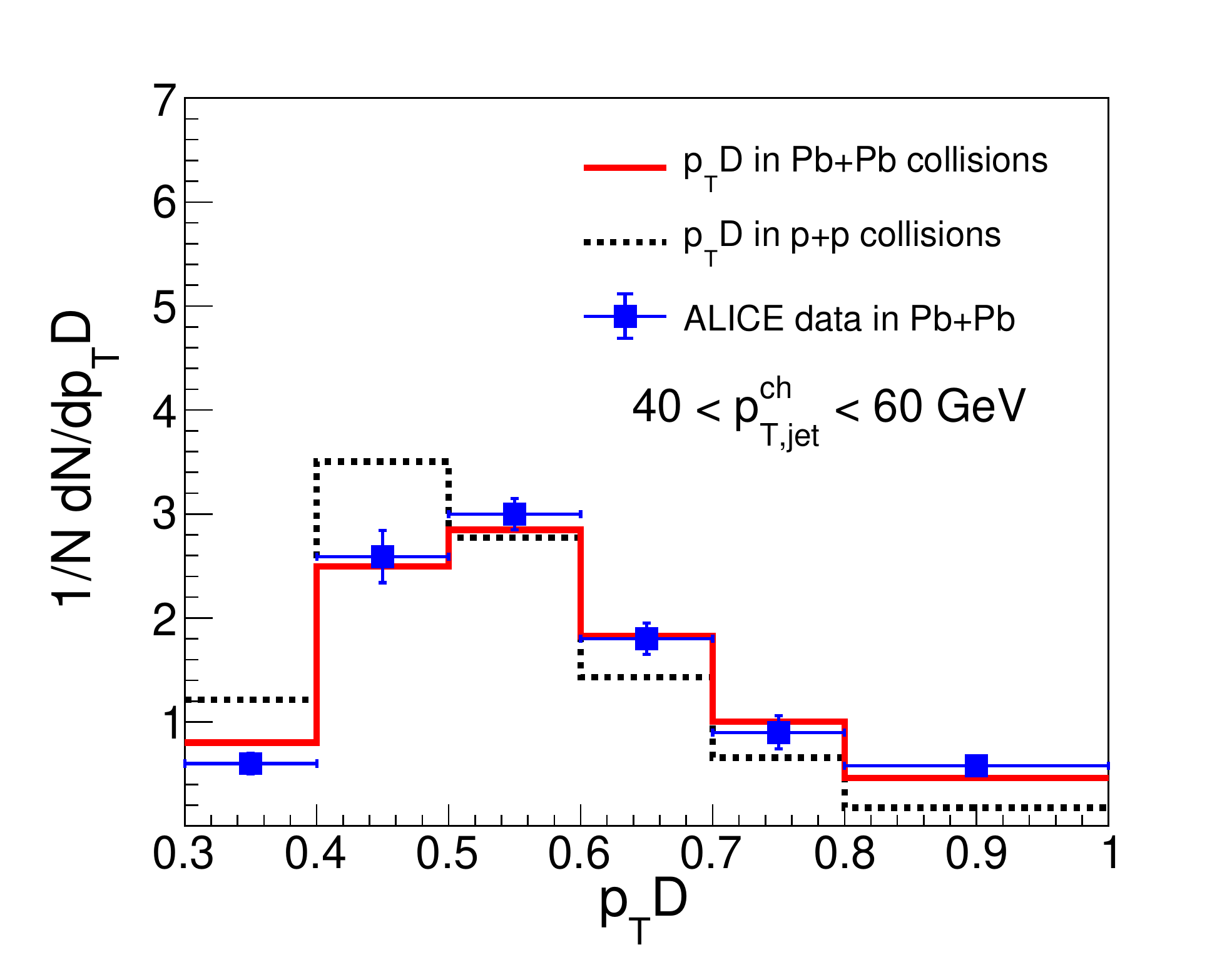} \\
\caption{Normalized $p_{T}D$ distributions of inclusive jets in p+p and 0-10\% Pb+Pb collisions at $\sqrt{s_{NN}}=2.76$~TeV as compared with ALICE data~\cite{Acharya:2018uvf}.}
\label{g-AA}
\end{figure}

\section{Results and discussion}

\subsection{$p_{T}D$ distributions of ungroomed jets in Pb+Pb collisions}

We now could calculate
the jet number normalized $p_{T}D$ distributions in Pb+Pb collisions at $\sqrt{s_{NN}}=2.76$~TeV.
We use the same jet selection criterion as we did in p+p collisions in Sec.~\ref{sec:framework}.
Our numerical results of jet number normalized $p_{T}D$ distributions for inclusive jets in p+p and Pb+Pb collisions
are shown in Fig.~\ref{g-AA}, which are confronted with the existing
experimental data in Pb+Pb by ALICE Collaboration~\cite{Acharya:2018uvf}. We find our theoretical calculations could provide quite decent descriptions of experimental measurements.
Relative to that in p+p collisions, the observed normalized $p_{T}D$ distribution in Pb+Pb is shifted towards higher values.
That is to say, $p_{T}D$ distribution of inclusive jets in Pb+Pb collisions is shifted towards quark jets after jet quenching.
It indicate that a jet in Pb+Pb collisions may have more softer constituents than that in p+p.

\begin{figure}[!htb]
\centering
\includegraphics[width=9.5cm,height=9.6cm]{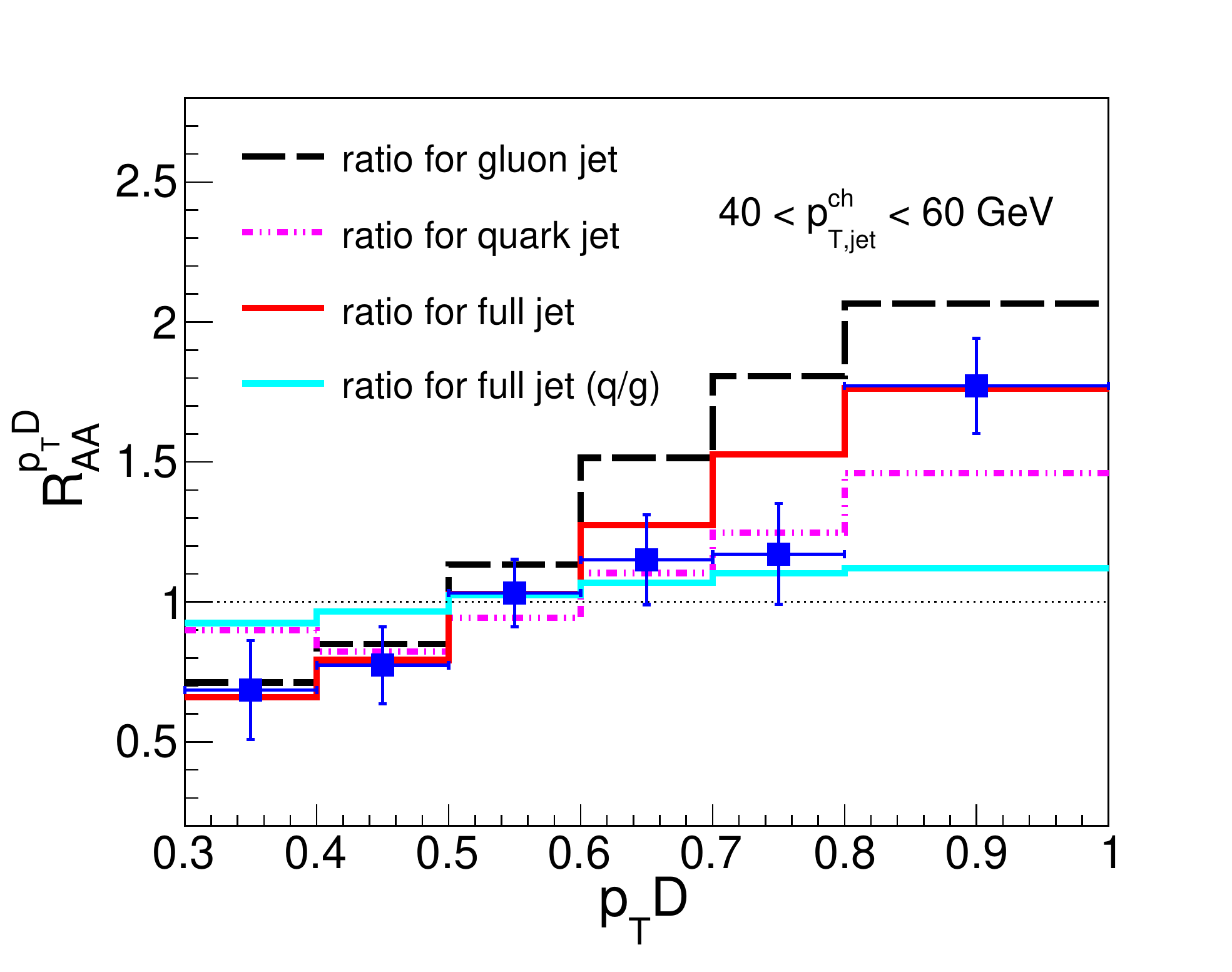} \\
\caption{Nuclear modification ratio of $p_{T}D$ distribution of inclusive jets as well as quark and gluon jets. The ratio from ALICE Collaboration is preformed by Pb+Pb measurements scaled by MC simulation in p+p~\cite{Acharya:2018uvf}.
}
\label{ratio-jets}
\end{figure}

To investigate the deviation of jet $p_{T}D$ distributions in HIC to those in p+p in a more straightforward way,
it is essential and helpful to define
the nuclear modification ratio($R_{AA}^{\rm p_{T}D}$) of $p_{T}D$ distributions as:
\begin{eqnarray}
R_{AA}^{\rm p_{T}D}={\dfrac{1}{N_{AA}} \dfrac{dN_{AA}}{dp_{T}D}}/{\dfrac{1}{N_{pp}} \dfrac{dN_{pp}}{dp_{T}D}} \,\, .
\label{eq:ratio}
\end{eqnarray}

Shown in Fig.~\ref{ratio-jets} are $R_{AA}^{\rm p_{T}D}$ of normalized $p_{T}D$ distributions for inclusive jets, as compared with ALICE ratio~\cite{Acharya:2018uvf}. Our calculated results could provide nice description of ALICE ratio in the overall $p_{T}D$ region. We note that, the ALICE ratio is preformed form Pb+Pb
measurements scaled by MC simulation in p+p due to the lack of corresponding measurement in p+p.
It is a normal practice to utilize MC simulations in p+p as reference in the studying of
nuclear modification ratios when the corresponding p+p baseline is not available~\cite{ALICE:2017nij,ALICE:2018lyv,ALICE:2018hbc,ALICE:2021vrw}. Of course, truly measured results of $R_{AA}$ in experiments are more favorable, and then we could confront our simulations directly with the experimental data. The $R_{AA}^{\rm p_{T}D}$ of quark jets and gluon jets as the components of inclusive jets are also plotted in Fig.~\ref{ratio-jets}. One can observe there is a suppression of $p_{T}D$ distribution for both quark jets and gluon jets at low $p_{T}D$ region,
while an enhancement at high $p_{T}D$ region. The nuclear corrections of gluon jets $p_{T}D$ distribution is much stronger than that for quark jets. The curve of $R_{AA}^{\rm p_{T}D}$ for inclusive jets goes between the curves of $R_{AA}^{\rm p_{T}D}$ for quark jets and gluon jets, since inclusive jets are the combinations of quark jets and gluon jets.

To further understand the nuclear modification mechanism of $p_{T}D$ distribution for inclusive jets, we start with the modifications of the relative fraction of quark and gluon jets due to jet quenching.
In our calculations, as most of other jet quenching models, gluons may lose more energy than quarks in QGP with their larger color charge. Therefore, generally we should see an enhancement of contribution fraction of quark jets in A+A collisions relative to that in p+p. Qualitatively such enhancement will naturally lead $p_{T}D$ distributions for inclusive jets to larger value region, since quark jets peak at larger value of $p_{T}D$ than gluon jets.
This is illustrated in Fig.~\ref{ratio-jets}, where the curve labelled ``inclusive jet (q/g)" represents our numerical result of $R_{AA}^{\rm p_{T}D}$ by only considering the effect of quark/gluon jets fraction alterations due to jet quenching while assuming there are no medium modifications for $p_{T}D$ distributions of pure quark and gluon jets in heavy-ion collisions.

\begin{figure}[!htb]
\centering
\includegraphics[width=9.5cm,height=9.6cm]{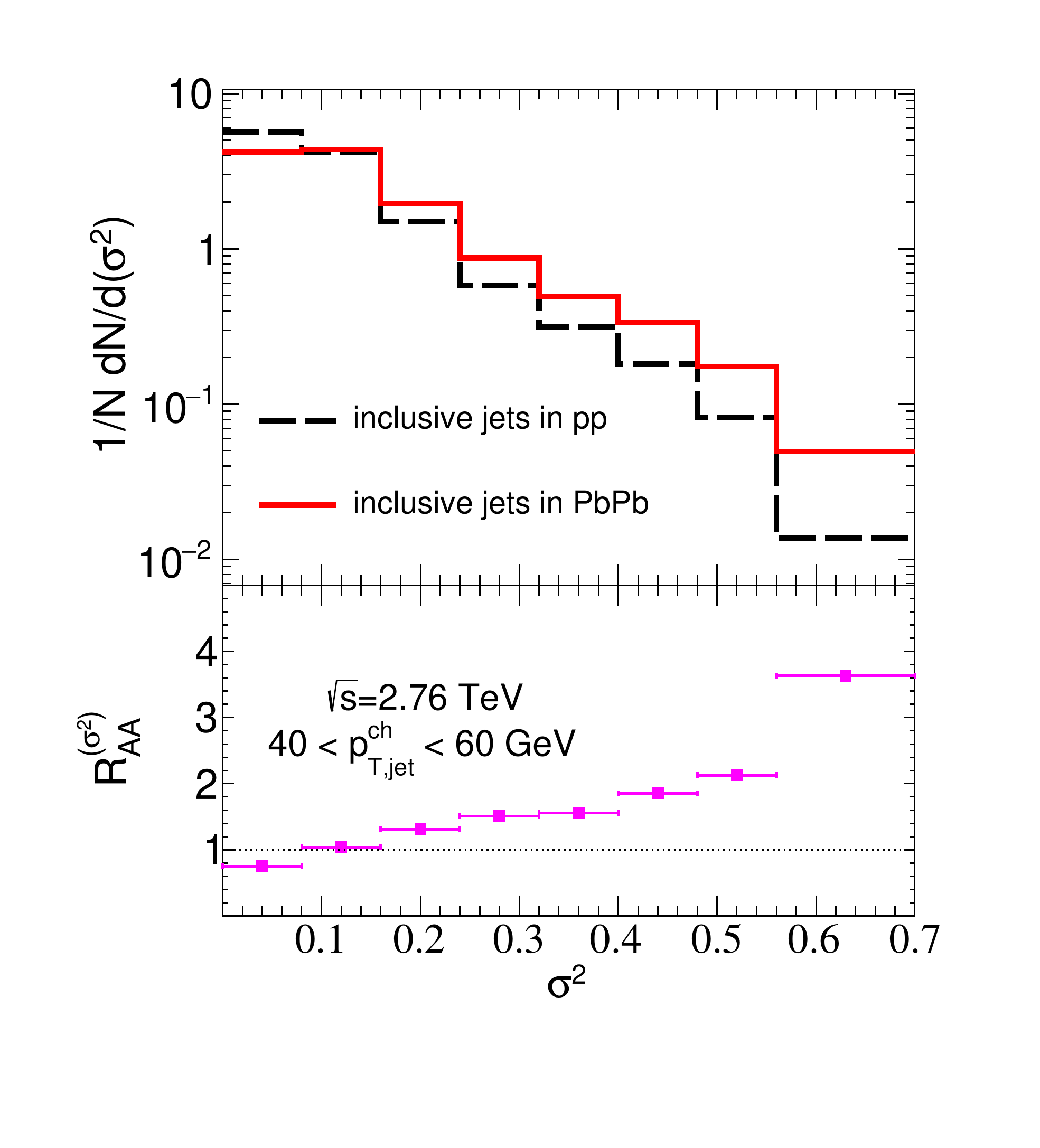} \\
\vspace{-1cm}
\caption{Top: normalized $\sigma^2$ distribution of inclusive jets in 0-10\% Pb+Pb and p+p collisions at
$\sqrt{s_{NN}}=2.76$~TeV; Bottom: ratio of normalized $\sigma^2$ distribution in central Pb+Pb (0-10\%) and p+p collisions.}
\label{delta2aa}
\end{figure}

\begin{figure}[!htb]
\centering
\vspace{0.in}
\includegraphics[width=9.5cm,height=9.6cm]{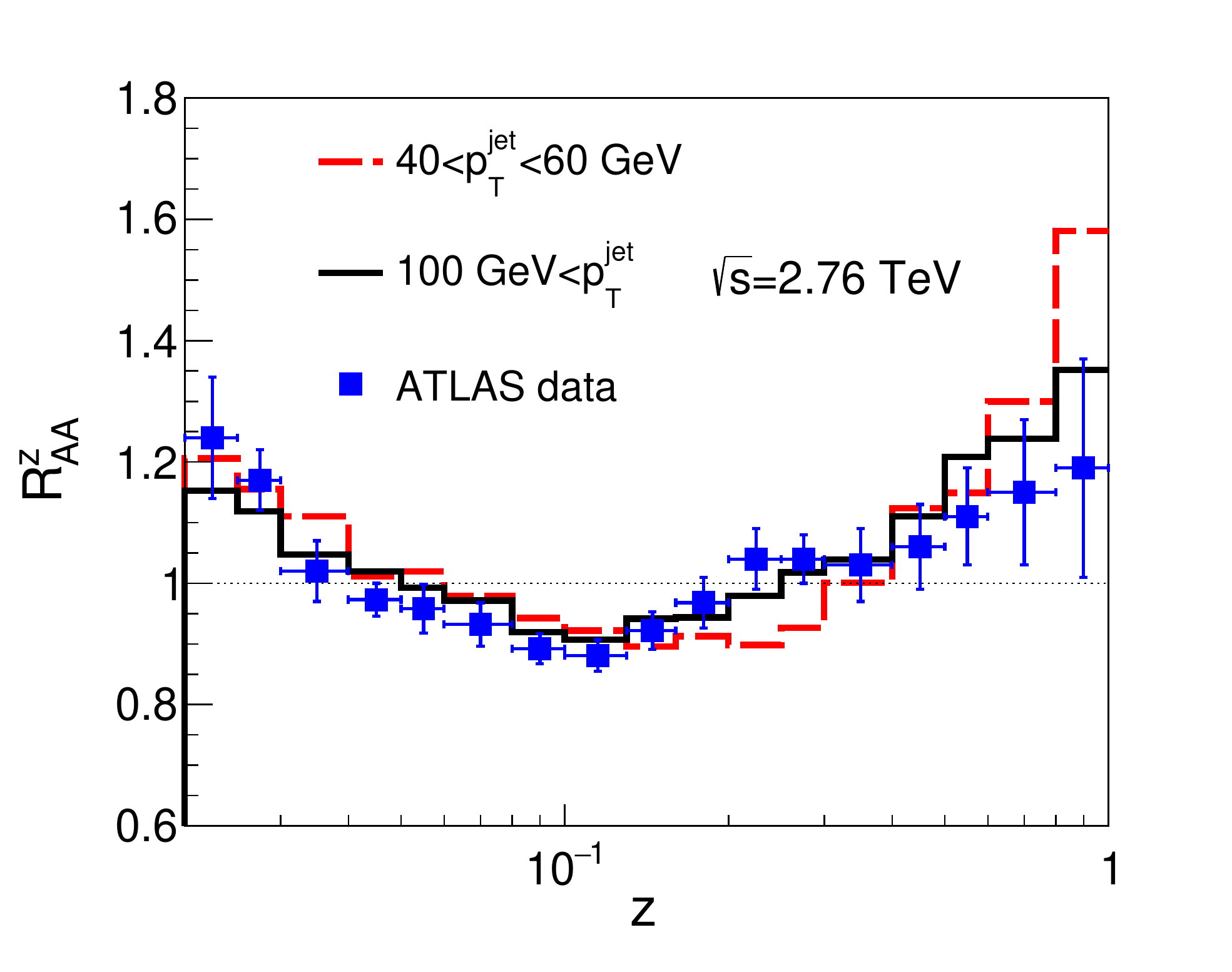} \\
\vspace{0.in}
\caption{Nuclear modification ratio of z distributions for inclusive jets at $\sqrt{s_{NN}}=2.76$~TeV as compared with ATLAS data~\cite{Aad:2014wha}.}
\label{zratio}
\end{figure}

To explore why for normalized jet $p_{T}D$ distributions $R_{AA}^{\rm p_{T}D} < 1$ at small $p_{T}D$ region, and $R_{AA}^{\rm p_{T}D} > 1$ at large $p_{T}D$ region, as shown in Fig.~\ref{ratio-jets}. Eq.(\ref{eq:sigma}) indicate the nuclear modification of $p_{T}D$ distributions has a very strong correlation with the nuclear correction of standard deviation of $p_{T,i}$.
Thus we turn to investigate the standard deviation of $p_{T,i}$ for jets in p+p and 0-10\% Pb+Pb collisions. Presented in Fig.~\ref{delta2aa} are the normalized distribution of variance ($\sigma^{2}$) of $p_{T,i}$ in p+p and 0-10\% Pb+Pb collisions at $\sqrt{s_{NN}}=2.76$~TeV. We can observed the distributions of jet variance is also shifted to higher value region in heavy-ion collisions compared with that in p+p. Which indicates after jet quenching, the value of $p_{T,i}$ lies further from the mean value in heavy-ion collisions relative to p+p. It is noted that the changes of mean multiplicities of Pb+Pb relative to p+p is rather small, and the estimated mean values of jet constituents number ($\bar{n}$) in p+p and Pb+Pb collisions are $\bar{n}_{pp}=6.72$ and $\bar{n}_{PbPb}=6.54$ respectively.

To see the point more clearly, we also plotted the nuclear modification ratio of momentum fraction for jet constituents($z=p_{T,i}/p_{T, \rm jet}$) in Fig.~\ref{zratio}. One can see, our model calculations could provide nice description of ATLAS data for jets with $p_T>100$~GeV~\cite{Aad:2014wha}. For jets with $40<p_T<60$~GeV, the nuclear modification ratios of charged-particle transverse momentum distributions in Pb+Pb collisions to those measured in p+p exhibit an enhancement in fragment yield in central collisions for $0.02<z<0.05$,
a reduction in fragment yields for $0.05<z<0.3$,
and an enhancement in the fragment yield for $0.3<z<1$. Which means the number of jet constituents with lower and higher value of $p_{T}$ are enhanced, more constituents lied further from the mean value. Therefore, the jet standard deviation is shifted to higher region in heavy-ion collisions.

\subsection{$p_{T}D$ distributions of groomed jets in Pb+Pb collisions}
In this section, we will study $p_{T}D$ distributions of groomed jets in central Pb+Pb collisions. Jet grooming techniques have seen a particularly great of interest from both experimental and theoretical side~\cite{Larkoski:2014wba,Dasgupta:2013ihk}. It is designed to remove soft wide-angle radiation from the jet,
allowing for a more direct comparison between experimental data and purely perturbative QCD calculations, since hadronization and underlying event contributions are significantly reduced during grooming procedure.
A full jet constructed using radius $R$ via the anti-$k_{T}$ algorithm is first re-clustered using the Cambridge-Aachen (C/A) algorithm~\cite{Dokshitzer:1997in,Wobisch:1998wt} until two hard branches are found to satisfy the following condition:

\begin{eqnarray}
\frac{min(p_{T1},p_{T2})}{p_{T1}+p_{T2}}\equiv{z_{g}}>z_{cut}\left (\frac{\Delta R}{R} \right)^{\beta}
\label{eq:sd}
\end{eqnarray}
where $\left (\frac{\Delta R}{R} \right)$ is an additional parameter of the relative angular distance between the two sub-jets, $z_{cut}$ and $\beta$ are free
parameters which can be used to control how strict the soft drop condition is. For the heavy-ion studies conducted so far, $z_{cut}$ has been set to $0.1$ and $\beta$
has been set to zero.

\begin{figure}[!htb]
\centering
\includegraphics[width=9.5cm,height=9.6cm]{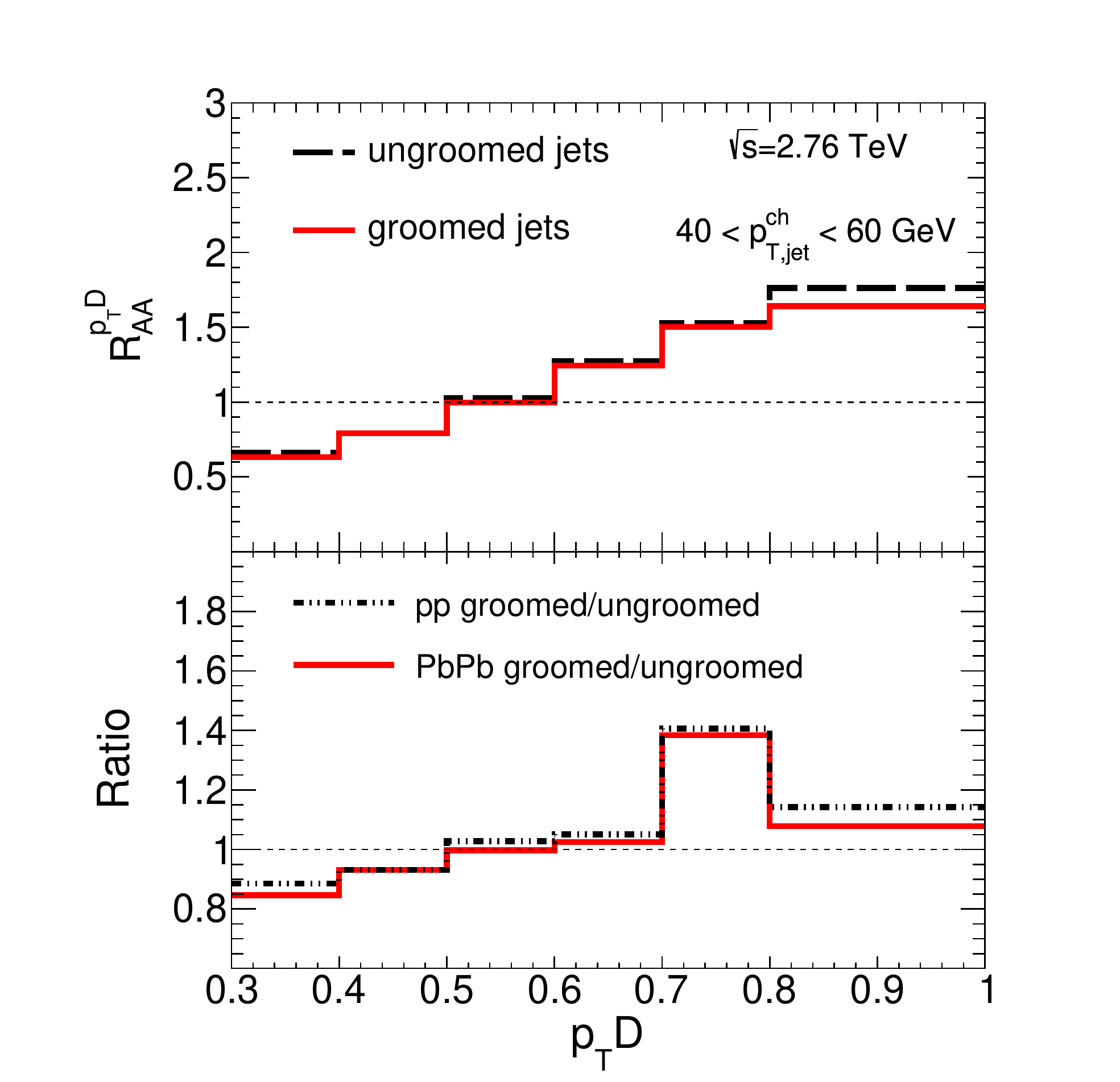} \\
\vspace{0.2cm}
\caption{Top: nuclear modification factor of $p_{T}D$ distribution for groomed and ungroomed jets; Bottom: ratio of $p_{T}D$ distribution for groomed and ungroomed jets in p+p and 0-10\% Pb+Pb collisions.}
\label{delta2aag}
\end{figure}

In the top panel of Fig.~\ref{delta2aag}, we plot the nuclear modification ratio of $p_{T}D$ distributions for groomed and ungroomed jets respectively. One can observe that the nuclear modification pattern
of $p_{T}D$ distributions for groomed jets are similar with that for ungroomed jets, and
the $p_{T}D$ distribution for groomed jets
is also shifted to higher $p_{T}D$ region. Besides, compared with ungroomed jets,
the nuclear modification of groomed jets becomes weaker. It implies grooming procedure could not only remove soft radiation from the jet in QCD vacuum, but also reduce soft radiation
in QCD medium. In the bottom panel of Fig.~\ref{delta2aag} we present the ratios of $p_{T}D$ distributions for the groomed jet
to that for
the ungroomed both in p+p and Pb+Pb collisions. It is shown that these ratios are below to unity at small $p_{T}D$, whereas
larger than one at large $p_{T}D$. To understand deeper the alteration of $p_{T}D$ distributions originate from jet grooming procedure, in the following we may investigate the difference of standard deviation of jet constituents $p_{T,i}$ and jet constituents number between groomed and ungroomed jets in Pb+Pb collisions.

\begin{figure}[!htb]
\centering
\includegraphics[width=9.5cm,height=9.6cm]{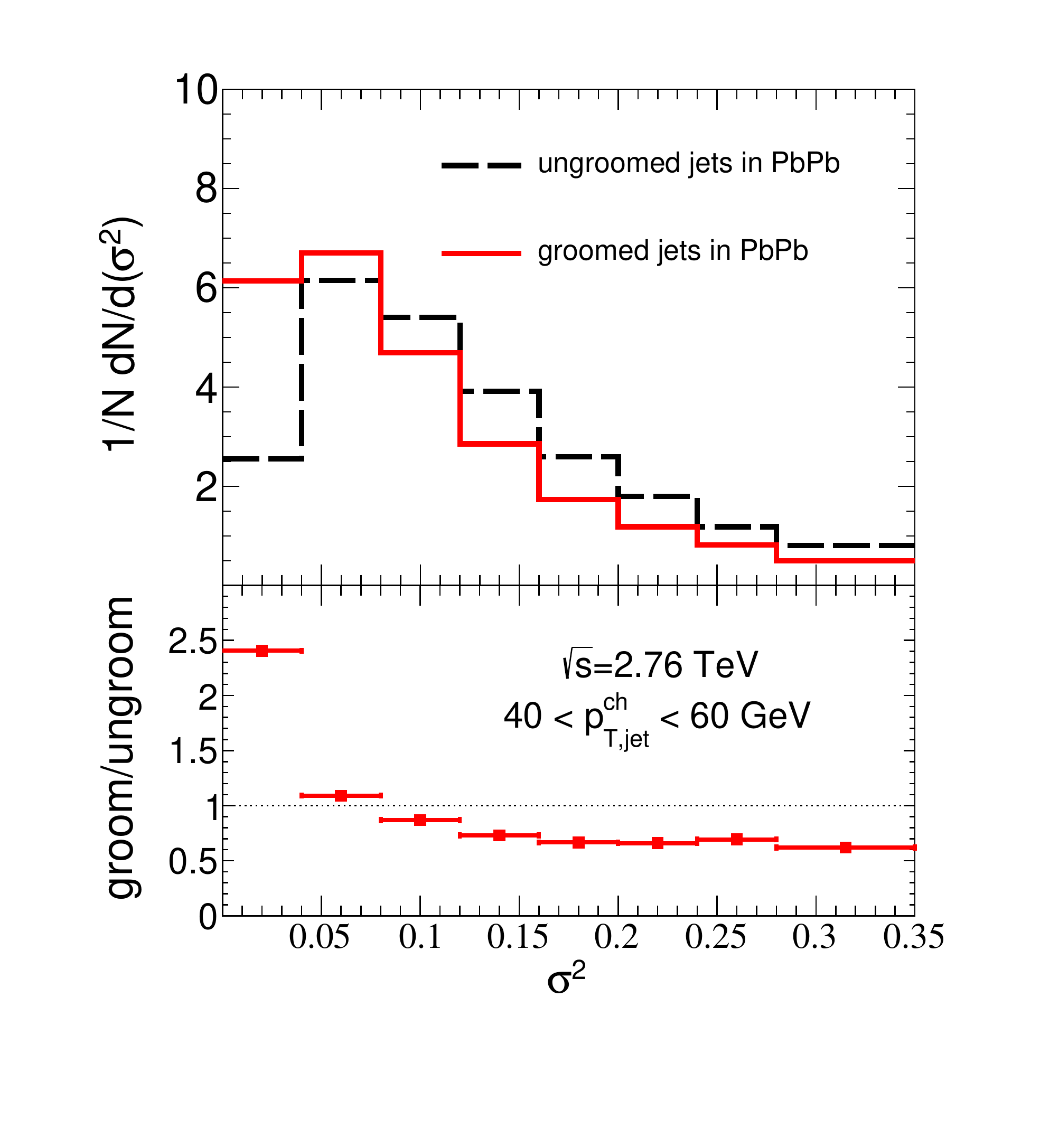} \\
\vspace{-1cm}
\caption{Top: normalized $\sigma^2$ distribution of groomed and ungroomed jets in 0-10\% Pb+Pb collisions at
$\sqrt{s_{NN}}=2.76$~TeV; Bottom: ratio of normalized $\sigma^2$ distribution for groomed and ungroomed jets in 0-10\% Pb+Pb collisions.}
\label{delta2aa2}
\end{figure}

Firstly, in Fig.~\ref{delta2aa2} we plotted the variance ($\sigma^{2}$) distributions of jet constituents $p_{T,i}$ for groomed and ungroomed jets  in central Pb+Pb collisions. The distributions of jet $\sigma^{2}$ are shifted to lower region for groomed jets compared with that for ungroomed jets, which is in contrast to the alteration of $p_{T}D$ distributions. It indicates that after jet soft-drop procedure, the values of $p_{T,i}$ in groomed jets lies closer to the mean value relative to those in ungroomed jets. That is because during the grooming, some particles with low $p_{T,i}$ are dropped from the jet constituents.

\begin{figure}[!htb]
\includegraphics[width=9.5cm,height=9.6cm]{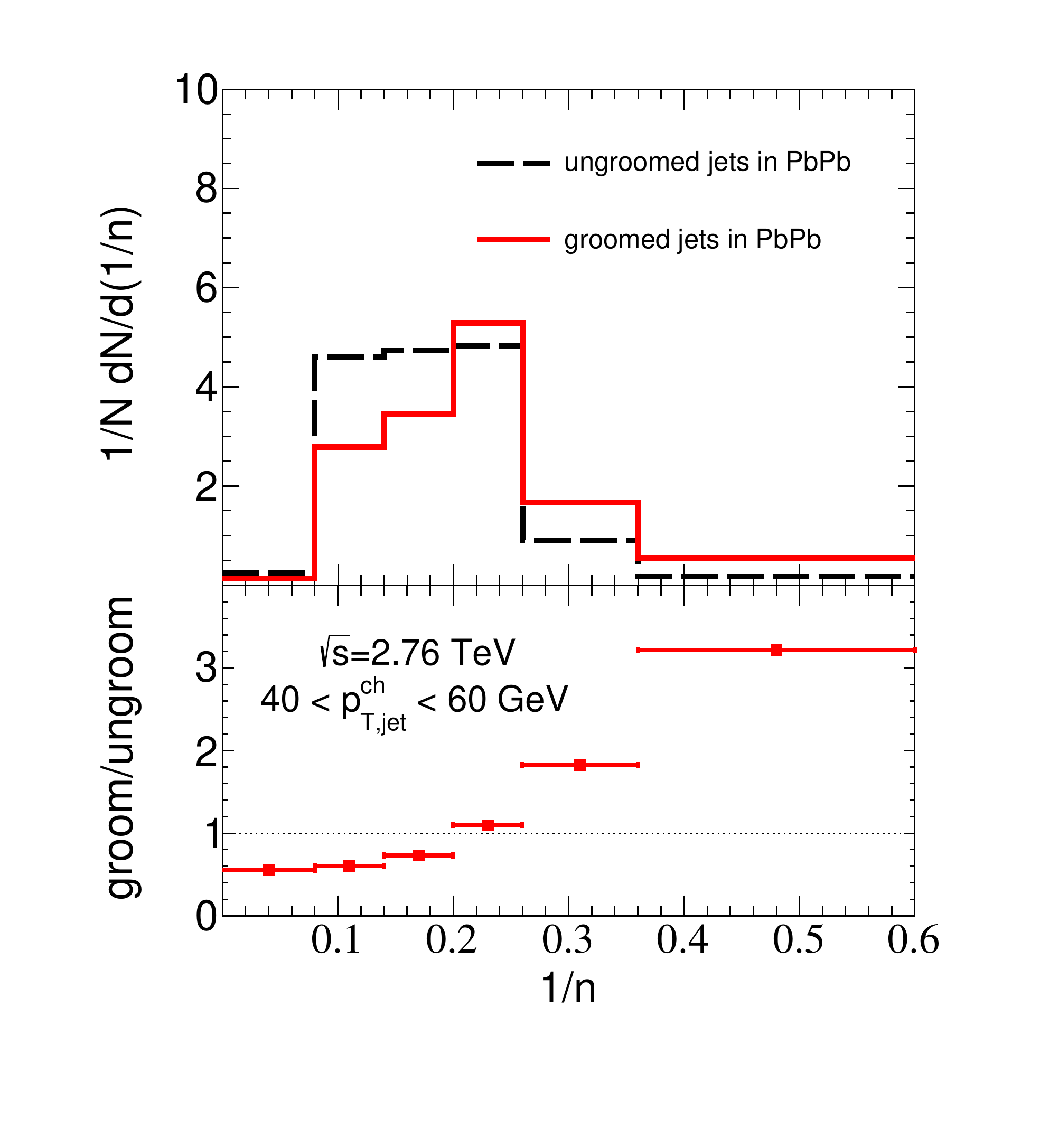} \\
\vspace{-1cm}
\caption{Top: normalized $(1/n)$ distribution of groomed and ungroomed jets in  0-10\% Pb+Pb collisions at
$\sqrt{s_{NN}}=2.76$~TeV; Bottom: ratio of normalized $(1/n)$ distribution for groomed and ungroomed jets in  0-10\% Pb+Pb collisions.}
\label{nn2aa}
\end{figure}

Secondly, the number of jets constituents are modified during soft drop grooming process, which will be contribute to the correction of $p_{T}D$ distributions. Presented in Fig.~\ref{nn2aa} are the number of jets constituents distributions of groomed and ungroomed jets in p+p and central Pb+Pb collisions. One cam see the value of $1/n$ is enhanced after grooming process. As we shown in Eq.~3, the value of $p_{T}D$ is equivalent to the standard deviation added by $1/n$. Therefore, even though the grooming process will lead to lower value of standard deviation for jet constituents transverse momenta, it will enhance the value of $1/n$ meanwhile. Though these two effects offset with each other, the correction of $1/n$ are more pronounced, which results in the increase of the ratio of $p_{T}D$ distributions for groomed jets  to that for ungroomed jets with $p_{T}D$ (as shown in the bottom panel
of Fig.~\ref{delta2aag}), a trend similar to the increase of ratio of normalized ($1/n$) distribution for groomed to ungroomed jets with $1/n$ (as demonstrated in the bottom panel of Fig.~\ref{nn2aa}).

\section{Summary}
\label{sec:summary}
In this paper, by using a NLO$+$PS event generator POWHEG$+$PYTHIA for p+p baseline and HT parton energy loss approach for jet quenching, we have studied
the nuclear modifications of $p_{T}D$ distributions for inclusive jets with small radius $R=0.2$ in 0-10\% Pb+Pb at $\sqrt{s_{NN}}=2.76$~TeV.
Our simulated results of inclusive jets could provide decent description of ALICE measurements. The $p_{T}D$ distributions for inclusive jets
are shifted toward higher $p_{T}D$ region in central Pb+Pb collisions compared to those in p+p, and similar trends have also been found for quark and gluon jets.
We further find two elements could contribute to the nuclear modifications of $p_{T}D$ distributions: more uneven $p_T$ of jet constituents, and the enhanced fraction of quark-initiated jets after jet-medium interaction in HIC.
The observed nuclear modifications of $p_{T}D$ distributions for gluon jets are stronger than that for quark jets in HIC
since gluons may lose more energy than quarks in our model.
Additionally, we also investigate the medium modifications of $p_{T}D$ distributions for groomed jets in central Pb+Pb collisions. We observe weaker nuclear modifications of $p_{T}D$ distributions for groomed jets
compare to that for ungroomed jets.

{\bf Acknowledgments:}  The authors would like to thank P Ru, S L Zhang, S Wang and Q Zhang  for helpful discussions.This research was supported in
part by Guangdong Major Project of Basic and Applied Basic Research No. 2020B0301030008, by Natural Science Foundation of China (NSFC) under Project
Nos. 11935007 and and 12035007.  S.-Y. Chen is supported by the MOE Key Laboratory of Quark and Lepton Physics (CCNU) under Project
No. QLPL2020P01.


\vspace*{-.6cm}



\begin{thebibliography}{99}

\bibitem{Wang:1991xy}
  X.~N.~Wang and M.~Gyulassy,
  Phys.\ Rev.\ Lett.\  {\bf 68}, 1480 (1992).

\bibitem{Gyulassy:2003mc}
  M.~Gyulassy, I.~Vitev, X.~N.~Wang and B.~W.~Zhang,
  In *Hwa, R.C. (ed.) et al.: Quark gluon plasma* 123-191
  [nucl-th/0302077].

\bibitem{Qin:2015srf}
  G.~Y.~Qin and X.~N.~Wang,
  Int.\ J.\ Mod.\ Phys.\ E {\bf 24}, no. 11, 1530014 (2015).

\bibitem{Khachatryan:2016odn}
CMS, V.~Khachatryan {\em et~al.},
\newblock JHEP {\bf 04}, 039 (2017), arXiv:1611.01664.

\bibitem{Acharya:2018qsh}
ALICE, S.~Acharya {\em et~al.},
\newblock JHEP {\bf 11}, 013 (2018), arXiv:1802.09145.

\bibitem{Aad:2015wga}
ATLAS, G.~Aad {\em et~al.},
\newblock JHEP {\bf 09}, 050 (2015), arXiv:1504.04337.

\bibitem{Burke:2013yra}
JET, K.~M. Burke {\em et~al.},
\newblock Phys. Rev. {\bf C90}, 014909 (2014), arXiv:1312.5003.



\bibitem{Chen:2010te}
  X.~Chen, C.~Greiner, E.~Wang, X.~N.~Wang and Z.~Xu,
  Phys.\ Rev.\ C {\bf 81}, 064908 (2010).

\bibitem{Chen:2011vt}
  X.~Chen, T.~Hirano, E.~Wang, X.~N.~Wang and H.~Zhang,
  Phys.\ Rev.\ C {\bf 84}, 034902 (2011).

\bibitem{Liu:2015vna}
Z.~Liu, H.~Zhang, B.~Zhang and E.~Wang,
Eur. Phys. J. C \textbf{76}, no.1, 20 (2016)
[arXiv:1506.02840 [nucl-th]].

\bibitem{Dai:2015dxa}
W.~Dai, X.~Chen, B.~Zhang and E.~Wang,
Phys. Lett. B \textbf{750}, 390-395 (2015)
[arXiv:1506.00838 [nucl-th]].

\bibitem{Dai:2017piq}
W.~Dai, X.~Chen, B.~Zhang, H.~Zhang and E.~Wang,
Eur. Phys. J. C \textbf{77}, no.8, 571 (2017)
[arXiv:1702.01614 [nucl-th]].


\bibitem{Dai:2017tuy}
  W.~Dai, B.~W.~Zhang and E.~Wang,
   Phys.\ Rev.\ C {\bf 98}, 024901 (2018).



\bibitem{Ma:2018swx}
  G.~Y.~Ma, W.~Dai, B.~W.~Zhang and E.~K.~Wang,
  Eur.\ Phys.\ J.\ C {\bf 79}, no. 6, 518 (2019).


\bibitem{Xie:2019oxg}
  M.~Xie, S.~Y.~Wei, G.~Y.~Qin and H.~Z.~Zhang,
  Eur.\ Phys.\ J.\ C {\bf 79}, no. 7, 589 (2019).

\bibitem{Zhang:2022fau}
Q.~Zhang, W.~Dai, L.~Wang, B.~W.~Zhang and E.~Wang,
[arXiv:2203.10742 [hep-ph]].

\bibitem{Aad:2010bu}
Atlas Collaboration, G.~Aad {\em et~al.},
\newblock Phys.Rev.Lett. {\bf 105}, 252303 (2010), arXiv:1011.6182.

\bibitem{Chatrchyan:2011sx}
CMS, S.~Chatrchyan {\em et~al.},
\newblock Phys. Rev. {\bf C84}, 024906 (2011), arXiv:1102.1957.

\bibitem{Chatrchyan:2012gt}
CMS, S.~Chatrchyan {\em et~al.},
\newblock Phys. Lett. {\bf B718}, 773 (2013), arXiv:1205.0206.

\bibitem{Aad:2014bxa}
ATLAS, G.~Aad {\em et~al.},
\newblock Phys. Rev. Lett. {\bf 114}, 072302 (2015), arXiv:1411.2357.

\bibitem{Chatrchyan:2012gw}
CMS Collaboration, S.~Chatrchyan {\em et~al.},
\newblock JHEP {\bf 1210}, 087 (2012), arXiv:1205.5872.

\bibitem{Chatrchyan:2013kwa}
CMS Collaboration, S.~Chatrchyan {\em et~al.},
\newblock Phys.Lett. {\bf B730}, 243 (2014), arXiv:1310.0878.

\bibitem{Aad:2014wha}
ATLAS, G.~Aad {\em et~al.},
\newblock Phys. Lett. {\bf B739}, 320 (2014), arXiv:1406.2979.

\bibitem{Sirunyan:2017jic}
A.~M.~Sirunyan \textit{et al.} [CMS],
Phys. Rev. Lett. \textbf{119}, no.8, 082301 (2017)
[arXiv:1702.01060 [nucl-ex]].

\bibitem{Sirunyan:2018ncy}
A.~M.~Sirunyan \textit{et al.} [CMS],
Phys. Rev. Lett. \textbf{122}, no.15, 152001 (2019)
[arXiv:1809.08602 [hep-ex]].


\bibitem{Vitev:2008rz}
  I.~Vitev, S.~Wicks and B.~W.~Zhang,
  JHEP {\bf 0811}, 093 (2008).

\bibitem{Vitev:2009rd}
  I.~Vitev and B.~W.~Zhang,
  Phys.\ Rev.\ Lett.\  {\bf 104}, 132001 (2010).

\bibitem{CasalderreySolana:2010eh}
  J.~Casalderrey-Solana, J.~G.~Milhano and U.~A.~Wiedemann,
  J.\ Phys.\ G {\bf 38}, 035006 (2011).

\bibitem{Young:2011qx}
  C.~Young, B.~Schenke, S.~Jeon and C.~Gale,
  Phys.\ Rev.\ C {\bf 84}, 024907 (2011).

\bibitem{He:2011pd}
  Y.~He, I.~Vitev and B.~W.~Zhang,
  Phys.\ Lett.\ B {\bf 713}, 224 (2012).


 \bibitem{ColemanSmith:2012vr}
  C.~E.~Coleman-Smith and B.~Muller,
  Phys.\ Rev.\ C {\bf 86}, 054901 (2012).

\bibitem{Neufeld:2010fj}
  R.~B.~Neufeld, I.~Vitev and B.-W.~Zhang,
  Phys.\ Rev.\ C {\bf 83}, 034902 (2011).

\bibitem{Zapp:2012ak}
  K.~C.~Zapp, F.~Krauss and U.~A.~Wiedemann,
  JHEP {\bf 1303}, 080 (2013).

\bibitem{Dai:2012am}
  W.~Dai, I.~Vitev and B.~W.~Zhang,
  Phys.\ Rev.\ Lett.\  {\bf 110}, 142001 (2013).


\bibitem{Ma:2013pha}
  G.~L.~Ma,
  Phys.\ Rev.\ C {\bf 87}, no. 6, 064901 (2013).


\bibitem{Senzel:2013dta}
  F.~Senzel, O.~Fochler, J.~Uphoff, Z.~Xu and C.~Greiner,
  J.\ Phys.\ G {\bf 42}, no. 11, 115104 (2015).

\bibitem{Casalderrey-Solana:2014bpa}
J.~Casalderrey-Solana, D.~C. Gulhan, J.~G. Milhano, D.~Pablos, and
  K.~Rajagopal,
\newblock JHEP {\bf 10}, 019 (2014), arXiv:1405.3864,
\newblock [Erratum: JHEP09,175(2015)].

\bibitem{Milhano:2015mng}
  J.~G.~Milhano and K.~C.~Zapp,
  Eur.\ Phys.\ J.\ C {\bf 76}, no. 5, 288 (2016).

\bibitem{Chang:2016gjp}
  N.~B.~Chang and G.~Y.~Qin,
  Phys.\ Rev.\ C {\bf 94}, no. 2, 024902 (2016).


\bibitem{Majumder:2014gda}
  A.~Majumder and J.~Putschke,
  Phys.\ Rev.\ C {\bf 93}, no. 5, 054909 (2016).




\bibitem{Chen:2016cof}
  L.~Chen, G.~Y.~Qin, S.~Y.~Wei, B.~W.~Xiao and H.~Z.~Zhang,
  Phys.\ Lett.\ B {\bf 782}, 773 (2018).


\bibitem{Chien:2016led}
  Y.~T.~Chien and I.~Vitev,
  Phys.\ Rev.\ Lett.\  {\bf 119}, no. 11, 112301 (2017).

\bibitem{Apolinario:2017qay}
  L.~Apolinario, J.~G.~Milhano, M.~Ploskon and X.~Zhang,
  Eur.\ Phys.\ J.\ C {\bf 78}, no. 6, 529 (2018)


\bibitem{Connors:2017ptx}
  M.~Connors, C.~Nattrass, R.~Reed and S.~Salur,
  Rev.\ Mod.\ Phys.\  {\bf 90}, 025005 (2018)

\bibitem{Zhang:2018urd}
  S.~L.~Zhang, T.~Luo, X.~N.~Wang and B.~W.~Zhang,
  Phys.\ Rev.\ C {\bf 98}, 021901 (2018).

\bibitem{Dai:2018mhw}
W.~Dai, S.~Wang, S.~L.~Zhang, B.~W.~Zhang and E.~Wang,
Chin. Phys. C \textbf{44}, 104105 (2020)
[arXiv:1806.06332 [nucl-th]].

\bibitem{Luo:2018pto}
T.~Luo, S.~Cao, Y.~He and X.~Wang,
Phys. Lett. B \textbf{782}, 707-716 (2018)
[arXiv:1803.06785 [hep-ph]].

\bibitem{Chang:2019sae}
N.~Chang, Y.~Tachibana and G.~Qin,
Phys. Lett. B \textbf{801}, 135181 (2020)
[arXiv:1906.09562 [nucl-th]].

\bibitem{Wang:2019xey}
S.~Wang, W.~Dai, B.~W.~Zhang and E.~Wang,
Eur. Phys. J. C \textbf{79}, no.9, 789 (2019)
[arXiv:1906.01499 [nucl-th]].


\bibitem{Chen:2019gqo}
S.~Chen, B.~W.~Zhang and E.~Wang,
Chin. Phys. C \textbf{44}, no.2, 024103 (2020)
[arXiv:1908.01518 [nucl-th]].


\bibitem{Chen:2020kex}
L.~Chen, S.~Wei and H.~Zhang,
[arXiv:2001.07606 [hep-ph]].

\bibitem{Wang:2020qwe}
S.~Wang, W.~Dai, B.~W.~Zhang and E.~Wang,
[arXiv:2005.07018 [hep-ph]].

\bibitem{Yan:2020zrz}
J.~Yan, S.~Y.~Chen, W.~Dai, B.~W.~Zhang and E.~Wang,
Chin. Phys. C \textbf{45}, no.2, 024102 (2021)
[arXiv:2005.01093 [hep-ph]].

\bibitem{Wang:2020ukj}
S.~Wang, W.~Dai, B.~W.~Zhang and E.~Wang,
Chin. Phys. C \textbf{45}, no.6, 064105 (2021)
[arXiv:2012.13935 [nucl-th]].

\bibitem{Zhang:2021sua}
S.~L.~Zhang, M.~Q.~Yang and B.~W.~Zhang,
[arXiv:2105.04955 [hep-ph]].




\bibitem{Giele:1997hd}
  W.~T.~Giele, E.~W.~N.~Glover and D.~A.~Kosower,
  Phys.\ Rev.\ D {\bf 57}, 1878 (1998)
  [hep-ph/9706210].

\bibitem{Acharya:2018uvf}
  S.~Acharya {\it et al.} [ALICE Collaboration],
  JHEP {\bf 1810}, 139 (2018)
  [arXiv:1807.06854 [nucl-ex]].



\bibitem{KunnawalkamElayavalli:2017hxo}
R.~Kunnawalkam Elayavalli and K.~C.~Zapp,
JHEP \textbf{07}, 141 (2017)
[arXiv:1707.01539 [hep-ph]].

\bibitem{Agafonova:2019tqe}
  V.~Agafonova,
  Universe {\bf 5}, no. 5, 114 (2019).

\bibitem{Wan:2018zpq}
  R.~Z.~Wan, L.~Ding, X.~Gui, F.~Yang, S.~Li and D.~C.~Zhou,
  Chin.\ Phys.\ C {\bf 43}, no. 5, 054110 (2019)
  [arXiv:1812.10062 [hep-ph]].



\bibitem{Buckley:2016bhy}
  A.~Buckley and D.~Bakshi Gupta,
  arXiv:1608.03577 [hep-ph].


\bibitem{Alioli:2010qp}
S.~Alioli, P.~Nason, C.~Oleari and E.~Re,
JHEP \textbf{01}, 095 (2011)
[arXiv:1009.5594 [hep-ph]].


\bibitem{Alioli:2010xa}
S.~Alioli, K.~Hamilton, P.~Nason, C.~Oleari and E.~Re,
JHEP \textbf{04}, 081 (2011)
[arXiv:1012.3380 [hep-ph]].



\bibitem{Guo:2000nz}
  X.~F.~Guo and X.~N.~Wang,
  Phys.\ Rev.\ Lett.\  {\bf 85} (2000) 3591
  [hep-ph/0005044].

\bibitem{Zhang:2003yn}
  B.~W.~Zhang and X.~N.~Wang,
  Nucl.\ Phys.\ A {\bf 720}, 429 (2003).

\bibitem{Zhang:2003wk}
  B.~W.~Zhang, E.~Wang and X.~N.~Wang,
  Phys.\ Rev.\ Lett.\  {\bf 93} (2004) 072301
  [nucl-th/0309040].

\bibitem{Majumder:2009ge}
  A.~Majumder,
  Phys.\ Rev.\ D {\bf 85} (2012) 014023




\bibitem{Larkoski:2014pca}
A.~J.~Larkoski, J.~Thaler and W.~J.~Waalewijn,
JHEP \textbf{11}, 129 (2014)
[arXiv:1408.3122 [hep-ph]].

\bibitem{ALICE:2021njq}
S.~Acharya \textit{et al.} [ALICE],
[arXiv:2107.11303 [nucl-ex]].





\bibitem{Frixione:2007vw}
  S.~Frixione, P.~Nason and C.~Oleari,
  JHEP {\bf 0711}, 070 (2007)
  [arXiv:0709.2092 [hep-ph]].






\bibitem{powheg-box}
For more processes, please check the website: http://powhegbox.mib.infn.it

\bibitem{Skands:2010ak}
P.~Z.~Skands,
Phys. Rev. D \textbf{82}, 074018 (2010)
[arXiv:1005.3457 [hep-ph]].


\bibitem{Sjostrand:2006za}
T.~Sjostrand, S.~Mrenna and P.~Z.~Skands,
JHEP \textbf{05}, 026 (2006)
[arXiv:hep-ph/0603175 [hep-ph]].

\bibitem{Cacciari:2008gp}
  M.~Cacciari, G.~P.~Salam and G.~Soyez,
  JHEP {\bf 0804}, 063 (2008).

\bibitem{Alver:2008aq}
B.~Alver, M.~Baker, C.~Loizides and P.~Steinberg,
[arXiv:0805.4411 [nucl-ex]].



\bibitem{He:2015pra}
Y.~He, T.~Luo, X.~Wang and Y.~Zhu,
Phys. Rev. C \textbf{91}, 054908 (2015)
[arXiv:1503.03313 [nucl-th]].

\bibitem{Cao:2016gvr}
S.~Cao, T.~Luo, G.~Qin and X.~Wang,
Phys. Rev. C \textbf{94}, no.1, 014909 (2016)
[arXiv:1605.06447 [nucl-th]].


\bibitem{Cao:2017hhk}
S.~Cao, T.~Luo, G.~Qin and X.~Wang,
Phys. Lett. B \textbf{777}, 255-259 (2018)
[arXiv:1703.00822 [nucl-th]].



\bibitem{Neufeld:2010xi}
  R.~B.~Neufeld,
  Phys.\ Rev.\ D {\bf 83} (2011) 065012


\bibitem{Shen:2014vra}
  C.~Shen, Z.~Qiu, H.~Song, J.~Bernhard, S.~Bass and U.~Heinz,
  Comput.\ Phys.\ Commun.\  {\bf 199} (2016) 61



\bibitem{Lokhtin:2000wm}
I.~Lokhtin and A.~Snigirev,
Eur. Phys. J. C \textbf{16}, 527-536 (2000)
[arXiv:hep-ph/0004176 [hep-ph]].

\bibitem{Lokhtin:2005px}
I.~Lokhtin and A.~Snigirev,
Eur. Phys. J. C \textbf{45}, 211-217 (2006)
[arXiv:hep-ph/0506189 [hep-ph]].

\bibitem{ALICE:2017nij}
S.~Acharya \textit{et al.} [ALICE],
Phys. Lett. B \textbf{776}, 249-264 (2018)
[arXiv:1702.00804 [nucl-ex]].

\bibitem{ALICE:2018lyv}
S.~Acharya \textit{et al.} [ALICE],
JHEP \textbf{10}, 174 (2018)
[arXiv:1804.09083 [nucl-ex]].

\bibitem{ALICE:2018hbc}
S.~Acharya \textit{et al.} [ALICE],
Phys. Lett. B \textbf{793}, 212-223 (2019)
[arXiv:1809.10922 [nucl-ex]].


\bibitem{ALICE:2021vrw}
S.~Acharya \textit{et al.} [ALICE],
JHEP \textbf{10}, 003 (2021)
[arXiv:2105.04936 [nucl-ex]].


\bibitem{Larkoski:2014wba}
A.~J.~Larkoski, S.~Marzani, G.~Soyez and J.~Thaler,
JHEP \textbf{05}, 146 (2014)
[arXiv:1402.2657 [hep-ph]].

\bibitem{Dasgupta:2013ihk}
M.~Dasgupta, A.~Fregoso, S.~Marzani and G.~P.~Salam,
JHEP \textbf{09}, 029 (2013)
[arXiv:1307.0007 [hep-ph]].

\bibitem{Dokshitzer:1997in}
Y.~L.~Dokshitzer, G.~D.~Leder, S.~Moretti and B.~R.~Webber,
JHEP \textbf{08}, 001 (1997)
[arXiv:hep-ph/9707323 [hep-ph]].

\bibitem{Wobisch:1998wt}
M.~Wobisch and T.~Wengler,
[arXiv:hep-ph/9907280 [hep-ph]].


\end{thebibliography}
\end{document}